\documentclass[twocolumn]{aastex631}

\usepackage{booktabs}

\begin{document}

\title{Uncovering a Massive z$\sim$7.7 Galaxy Hosting a Heavily Obscured Radio-Loud AGN Candidate in COSMOS-Web}

\author[0000-0003-3216-7190]{Erini Lambrides}\altaffiliation{NPP Fellow}
\affiliation{NASA-Goddard Space Flight Center, Code 662, Greenbelt, MD, 20771, USA}
\correspondingauthor{Erini Lambrides}
\email{erini.lambrides@nasa.gov}

\author[0000-0003-1564-3802]{Marco Chiaberge}
\affil{Space Telescope Science Institute, 3700 San Martin Drive
Baltimore, MD 21218, USA}
\affil{Department of Physics \& Astronomy, Johns Hopkins University, Bloomberg Center, 3400 N. Charles St., Baltimore, MD 21218, USA}

\author[0000-0002-7530-8857]{Arianna S. Long}\altaffiliation{NASA Hubble Fellow}
\affiliation{Department of Astronomy, The University of Texas at Austin, 2515 Speedway Blvd Stop C1400, Austin, TX 78712, USA}

\author[0000-0001-9773-7479]{Daizhong Liu}
\affiliation{Max-Planck-Institut f\"{u}r Astrophysik, Karl-Schwarzschild-Str. 1, D-85748 Garching, Germany}

\author[0000-0003-3596-8794]{Hollis B. Akins}
\affiliation{Department of Astronomy, The University of Texas at Austin, 2515
Speedway Blvd Stop C1400, Austin, TX 78712, USA}

\author[0000-0001-5655-1440]{Andrew F. Ptak}
\affiliation{NASA-Goddard Space Flight Center, Code 662, Greenbelt, MD, 20771, USA}

\author[0000-0001-6102-9526]{Irham Taufik Andika}
\affiliation{Technical University of Munich, TUM School of Natural Sciences, Department of Physics, James-Franck-Str. 1, D-85748 Garching, Germany}
\affiliation{Max-Planck-Institut f\"{u}r Astrophysik, Karl-Schwarzschild-Str. 1, D-85748 Garching, Germany}

\author[0000-0003-3684-4275]{Alessandro Capetti}
\affiliation{INAF \- Osservatorio Astrofisico di Torino, Strada Osservatorio 20, I\-10025 Pino Torinese, Italy}

\author[0000-0002-0930-6466]{Caitlin M. Casey}
\affiliation{Department of Astronomy, The University of Texas at Austin, 2515
Speedway Blvd Stop C1400, Austin, TX 78712, USA}
\affiliation{Cosmic Dawn Center (DAWN), Denmark}

\author[0000-0002-6184-9097]{Jaclyn B. Champagne}
\affiliation{Steward Observatory, University of Arizona, 933 N Cherry Ave, Tucson, AZ 85721, USA}

\author[0000-0003-4922-0613]{Katherine Chworowsky}
\altaffiliation{NSF Graduate Research Fellow}
\affiliation{Department of Astronomy, The University of Texas at Austin, 2515
Speedway Blvd Stop C1400, Austin, TX 78712, USA}

\author[0000-0001-6812-7938]{Tracy E. Clarke}
\affiliation{U. S. Naval Research Laboratory, 4555 Overlook Ave SW, Washington, DC, 20375, USA}

\author[0000-0003-3881-1397]{Olivia R. Cooper}\altaffiliation{NSF Graduate Research Fellow}
\affiliation{Department of Astronomy, The University of Texas at Austin, 2515
Speedway Blvd Stop C1400, Austin, TX 78712, USA}

\author[0000-0002-0786-7307]{Xuheng Ding}
\affiliation{Kavli Institute for the Physics and Mathematics of the Universe (Kavli IPMU, WPI), The University of Tokyo, Chiba 277-8583, Japan}

\author[0000-0001-9584-2531]{Dillon Z. Dong}
\affiliation{National Radio Astronomy Observatory, 1003 Lopezville Road, Socorro, NM, 87801}

\author[0000-0002-9382-9832]{Andreas L. Faisst}
\affiliation{Caltech/IPAC, MS 314-6, 1200 E. California Blvd. Pasadena, CA 91125, USA}

\author[0000-0002-2077-2046]{Jordan Y. Forman}
\affiliation{Southeastern Universities Research Association\\
Washington D.C.,}
\affiliation{NASA Goddard Space Flight Center\\
Greenbelt, MD 20771, USA}

\author[0000-0002-3560-8599]{Maximilien Franco}
\affiliation{Department of Astronomy, The University of Texas at Austin, 2515
Speedway Blvd Stop C1400, Austin, TX 78712, USA}

\author[0000-0001-9885-4589]{Steven Gillman}
\affiliation{Cosmic Dawn Center (DAWN), Denmark}
\affiliation{DTU-Space, Technical University of Denmark, Elektrovej 327, DK-2800 Kgs. Lyngby, Denmark}

\author[0000-0002-0236-919X]{Ghassem Gozaliasl}
\affiliation{Department of Computer Science, Aalto University, P. O. Box 15400, Espoo, FI-00076, Finland}
\affiliation{Department of Physics, University of Helsinki, P. O. Box 64, FI-00014, Helsinki, Finland}

\author[0000-0002-4176-845X]{Kirsten R. Hall}
\affiliation{Radio \& Geoastronomy Division, Center for Astrophysics $\vert$ Harvard \& Smithsonian, 60 Garden St. Cambridge, MA 02138, USA}

\author[0000-0003-0129-2079]{Santosh Harish}
\affiliation{Laboratory for Multiwavelength Astrophysics, School of Physics and Astronomy, Rochester Institute of Technology, 84 Lomb Memorial Drive, Rochester, NY 14623, USA}

\author[0000-0003-4073-3236]{Christopher C. Hayward}
\affiliation{Center for Computational Astrophysics, Flatiron Institute, 162 Fifth Avenue, New York, NY 10010, USA}

\author[0000-0002-3301-3321]{Michaela Hirschmann}
\affiliation{Institute of Physics, GalSpec, Ecole Polytechnique Federale de Lausanne, Observatoire de Sauverny, Chemin Pegasi 51, 1290 Versoix, Switzerland}
\affiliation{INAF, Astronomical Observatory of Trieste, Via Tiepolo 11, 34131 Trieste, Italy}

\author[0000-0001-6251-4988]{Taylor A. Hutchison}
\altaffiliation{NPP Fellow}
\affiliation{NASA-Goddard Space Flight Center, Code 662, Greenbelt, MD, 20771, USA}

\author[0000-0003-3804-2137]{Knud Jahnke}
\affiliation{Max Planck Institute for Astronomy, K\"onigstuhl 17, D-69117 Heidelberg, Germany}

\author[0000-0002-8412-7951]{Shuowen Jin}
\altaffiliation{Marie Curie Fellow}
\affiliation{Cosmic Dawn Center (DAWN), Denmark} 
\affiliation{DTU-Space, Technical University of Denmark, Elektrovej 327, 2800 Kgs. Lyngby, Denmark}

\author[0000-0001-9187-3605]{Jeyhan S. Kartaltepe}
\affiliation{Laboratory for Multiwavelength Astrophysics, School of Physics and Astronomy, Rochester Institute of Technology, 84 Lomb Memorial Drive, Rochester, NY 14623, USA}

\author[0009-0008-4614-5818]{Emma T. Kleiner}
\affiliation{Southeastern Universities Research Association\\
Washington D.C.,}
\affiliation{NASA Goddard Space Flight Center\\
Greenbelt, MD 20771, USA}

\author[0000-0002-6610-2048]{Anton M. Koekemoer}
\affiliation{Space Telescope Science Institute, 3700 San Martin Dr., Baltimore, MD 21218, USA}

\author[0000-0002-5588-9156]{Vasily Kokorev}
\affiliation{Kapteyn Astronomical Institute, University of Groningen, PO Box 800, 9700 AV Groningen, The Netherlands}

\author[0000-0003-0415-0121]{Sinclaire M. Manning}\altaffiliation{NASA Hubble Fellow}
\affiliation{Department of Astronomy, University of Massachusetts Amherst, MA 01003, USA}

\author[0000-0001-9189-7818]{Crystal L. Martin}
\affil{Department of Physics, University of California, Santa Barbara, Santa Barbara, CA 93109, USA}

\author[0000-0002-6149-8178]{Jed McKinney}
\affiliation{Department of Astronomy, The University of Texas at Austin, 2515
Speedway Blvd Stop C1400, Austin, TX 78712, USA}

\author[0000-0002-5222-5717]{Colin Norman}
\affil{Space Telescope Science Institute, 3700 San Martin Drive Baltimore, MD 21218, USA}
\affil{Department of Physics \& Astronomy, Johns Hopkins University, Bloomberg Center, 3400 N. Charles St., Baltimore, MD 21218, USA}

\author[0000-0003-1991-370X]{Kristina Nyland}
\affiliation{U.S. Naval Research Laboratory, 4555 Overlook Ave SW, Washington, DC 20375, USA}

\author[0000-0003-2984-6803]{Masafusa Onoue}
\affiliation{Kavli Institute for Astronomy and Astrophysics, Peking University, Beijing 100871, China}
\affiliation{Kavli Institute for the Physics and Mathematics of the Universe (Kavli IPMU, WPI), The University of Tokyo, Chiba 277-8583, Japan}

\author[0000-0002-4271-0364]{Brant E. Robertson}
\affiliation{Department of Astronomy and Astrophysics, University of California, Santa Cruz, 1156 High Street, Santa Cruz, CA 95064, USA}

\author[0000-0002-7087-0701]{Marko Shuntov}
\affiliation{Cosmic Dawn Center (DAWN), Copenhagen, Denmark}
\affiliation{Niels Bohr Institute, University of Copenhagen, Jagtvej 128, DK-2200, Copenhagen, Denmark}

\author[0000-0002-0000-6977]{John D. Silverman}
\affiliation{Kavli Institute for the Physics and Mathematics of the Universe (WPI), The University of Tokyo, Kashiwa, Chiba 277-8583, Japan}
\affiliation{Department of Astronomy, School of Science, The University of Tokyo, 7-3-1 Hongo, Bunkyo, Tokyo 113-0033, Japan}

\author[0000-0001-9935-6047]{Massimo Stiavelli}
\affil{Space Telescope Science Institute, 3700 San Martin Drive Baltimore, MD 21218, USA}

\author[0000-0002-3683-7297]{Benny Trakhtenbrot}
\affiliation{School of Physics and Astronomy, Tel Aviv University, Tel Aviv 69978, Israel}

\author[0000-0002-4437-1773]{Eleni Vardoulaki}
\affiliation{Th\"{u}ringer Landessternwarte, Sternwarte 5, 07778 Tautenburg, Germany}

\author[0000-0002-7051-1100]{Jorge A. Zavala}
\affiliation{National Astronomical Observatory of Japan, 2-21-1 Osawa, Mitaka, Tokyo 181-8588, Japan}

%%%%%%%%%%%%%%%%%%%%%%%%% COSMOS-Web architects %%%%%%%%%%%%%%%%%%%%%%%%%%
%%%% in alphabetical order 

\author[0000-0001-9610-7950]{Natalie Allen}
\affiliation{Cosmic Dawn Center (DAWN), Copenhagen, Denmark}
\affiliation{Niels Bohr Institute, University of Copenhagen, Jagtvej 128, DK-2200, Copenhagen, Denmark}

\author[0000-0002-7303-4397]{Olivier Ilbert}
\affiliation{Aix Marseille Univ, CNRS, CNES, LAM, Marseille, France  }

\author[0000-0002-9489-7765]{Henry Joy McCracken}
\affiliation{Institut d’Astrophysique de Paris, UMR 7095, CNRS, and Sorbonne Université, 98 bis boulevard Arago, F-75014 Paris, France}

\author[0000-0003-2397-0360]{Louise Paquereau} 
\affiliation{Institut d’Astrophysique de Paris, UMR 7095, CNRS, and Sorbonne Université, 98 bis boulevard Arago, F-75014 Paris, France}

\author[0000-0002-4485-8549]{Jason Rhodes}
\affiliation{Jet Propulsion Laboratory, California Institute of Technology, 4800 Oak Grove Drive, Pasadena, CA 91001, USA}

\author[0000-0003-3631-7176]{Sune Toft}
\affiliation{Cosmic Dawn Center (DAWN), Copenhagen, Denmark}
\affiliation{Niels Bohr Institute, University of Copenhagen, Jagtvej 128, DK-2200, Copenhagen, Denmark}

%% Note that the \and command from previous versions of AASTeX is now
%% depreciated in this version as it is no longer necessary. AASTeX 
%% automatically takes care of all commas and "and"s between authors names.

%% AASTeX 6.31 has the new \collaboration and \nocollaboration commands to
%% provide the collaboration status of a group of authors. These commands 
%% can be used either before or after the list of corresponding authors. The
%% argument for \collaboration is the collaboration identifier. Authors are
%% encouraged to surround collaboration identifiers with ()s. The 
%% \nocollaboration command takes no argument and exists to indicate that
%% the nearby authors are not part of surrounding collaborations.

%% Mark off the abstract in the ``abstract'' environment. 
\begin{abstract}

%, there is no direct detection of accretion disk emission in the rest-frame UV/optical. Despite the prediction that at z$>7$, $>99\%$ of powerfully accreting SMBHs (or QSOs) are heavily obscured

In this letter, we report the discovery of the highest redshift, heavily obscured, radio-loud AGN candidate selected using JWST NIRCam/MIRI, mid-IR, sub-mm, and radio imaging in the COSMOS-Web field. Using multi-frequency radio observations and mid-IR photometry, we identify a powerful, radio-loud (RL), growing supermassive black hole (SMBH) with significant spectral steepening of the radio SED ($f_{1.28 \mathrm{GHz}} \sim 2$ mJy, $q_{24\micron} = -1.1$, $\alpha_{1.28-3\mathrm{GHz}}=-1.2$, $\Delta \alpha = -0.4$). In conjunction with ALMA, deep ground-based observations, ancillary space-based data, and the unprecedented resolution and sensitivity of JWST, we find no evidence of AGN contribution to the UV/optical/NIR data and thus infer heavy amounts of obscuration (N$_{\mathrm{H}} > 10^{23}$ cm$^{-2}$). Using the wealth of deep UV to sub-mm photometric data, we report a singular solution photo-z of $z_\mathrm{phot}$ = 7.7$^{+0.4}_{-0.3}$ and estimate an extremely massive host-galaxy ($\log M_{\star} = 11.4 -12\,\mathrm{M}_{\odot}$) hosting a powerful, growing SMBH (L$_{\mathrm{Bol}} = 4-12 \times 10^{46}$ erg s$^{-1}$). This source represents the furthest known obscured RL AGN candidate, and its level of obscuration aligns with the most representative but observationally scarce population of AGN at these epochs. %The sub-mm properties of this source indicate it is 

%Additionally, this source, coupled with the existence of a previously identified z$_{\mathrm{spec}}= 6.8$ RL QSO within 1.5 sq. deg., yields a measured space density over 2000 times greater then predictions [insert number]. Thus, not only is this source the highest-z obscured RL AGN candidate reported, but it's existence provides significant challenges to current predictions on SMBH growth at early cosmic times.

\end{abstract}

%% Keywords should appear after the \end{abstract} command. 
%% The AAS Journals now uses Unified Astronomy Thesaurus concepts:
%% https://astrothesaurus.org
%% You will be asked to selected these concepts during the submission process
%% but this old "keyword" functionality is maintained in case authors want
%% to include these concepts in their preprints.
\keywords{}

%% From the front matter, we move on to the body of the paper.
%% Sections are demarcated by \section and \subsection, respectively.
%% Observe the use of the LaTeX \label
%% command after the \subsection to give a symbolic KEY to the
%% subsection for cross-referencing in a \ref command.
%% You can use LaTeX's \ref and \label commands to keep track of
%% cross-references to sections, equations, tables, and figures.
%% That way, if you change the order of any elements, LaTeX will
%% automatically renumber them.
%%
%% We recommend that authors also use the natbib \citep
%% and \citet commands to identify citations.  The citations are
%% tied to the reference list via symbolic KEYs. The KEY corresponds
%% to the KEY in the \bibitem in the reference list below. 

 \section{Introduction} \label{sec:intro}
Recent discoveries of $z>6$ extremely powerful ($L_\mathrm{Bol} \sim 10^{46}$\, erg\,s$^{-1}$) active galactic nuclei (hereinafter referred to as AGN) have launched intense debate as to how such massive black holes ($\sim$10$^{9}\,\mathrm{M}_{\odot}$) can form so early in the Universe \citep{Mort11, Banados18, Inay20, Wang21}. Questions surrounding the triggering and growth of these AGN have largely remained unanswered. This is driven by the fact that almost all direct observations of $z>6$ AGN are \textit{unobscured} -- the very energy that makes these sources detectable at high-redshifts overwhelms the star-forming (SF) contributions from their host galaxies in rest-frame UV--NIR imaging. 

Thus it is paramount to observe powerful AGN at $z>6$ whose central engines are heavily obscured for the following reasons: (1) Unlike with unobscured AGN, the host galaxy properties of obscured AGN (e.g., $M_{\star}$, morphology) are more accessible in regimes where the AGN emission is significantly attenuated (i.e., rest-frame UV/optical); (2) According to a combination of theory and observations over 80\% of AGN are expected to be heavily obscured (N$_{H} > 10^{23}$\,cm$^{-2}$) by their host-galaxies at $z > 6$, and over 99\% by $z>7$ \citep{ni2020,gilli22}. The obscuration of AGN can occur over a vast range of physical scales and conditions. In the local Universe, obscured AGN are contextualized by the standard sight-line dependent unification scheme -- where the dominant source of obscuration is thought to occur a few parsecs from the accretion disk by an optically thick toroidal or cloud structure and exhibit a lack of intrinsic difference between the host-galaxy and BH properties of their unobscured AGN counterparts \citep{antonuuci,urry95}. New evidence is accumulating that at higher redshifts, the dominant sources of AGN obscuration may exist on kpc scales \citep{circosta19,damato}. In both theory and observations, it is shown that at increasing redshifts, galaxies are clumpy and less settled \citep{Faure2021, Kartaltepe2023}, and have higher ISM densities \citep{Buchner2017, Dalton2021, gilli22}. Therefore, it is unsurprising that recent studies find high AGN obscured fractions due to the increased chances of UV/optical photons from the accretion disk being significantly attenuated along its path through its host galaxy \citep{ni2020,gilli22}. 

Recent JWST spectroscopic and photometric observations have yielded a litany of $z > 5$ actively accreting SMBHs \citep{kocevski23,larson23,labbe23,matthee,furtak,maiolino23a}, yet for these sources -- some of which are heavily reddened -- their rest-frame UV-Optical emission probes their AGN nature, and thus by definition are not heavily obscured. Even JWST/MIRI spectra of z $\sim$ 7 AGN probe rest-frame $\lesssim 2 \micron$ emission (i.e. \citealt{bosman23}), and for the most obscured AGN, their nature may only be robustly revealed at rest mid-infrared (MIR) wavelengths in lieu of sufficient detection of high-ionization lines \citep{hickox18}. Thus, these newly measured JWST sources may not represent the most common type of AGN at these epochs, and it is yet to be determined whether their formation and/or evolution is intrinsically different from the high-$z$ obscured AGN population. From black hole seeds to AGN feedback, the interpretation of JWST discovered high-$z$ AGN candidates may be significantly impacted if there are different triggering pathways or host-galaxy properties of obscured vs.\ unobscured AGN.

Despite the predicted increased number density of high-$z$ heavily obscured AGN, their identification is incredibly difficult due to their heavy obscuration at wavelengths shorter than the MIR and lack of observing facilities that are capable of probing the rest-frame MIR at these epochs. Rest-frame optical-NIR spectroscopy will lack the characteristic broad lines evident in unobscured sources and requires careful analysis of multiple, well-detected narrow lines to constrain whether the source of the ionizing photons is dominated by AGN vs.\ star-forming processes \citep{onoue}. In addition, X-ray facilities are generally incapable of reaching the sensitivities required for other than a handful of sources at $z=6$--7 \citep{vito2019} and a potentially lensed $z=10$ source \citep{akos, goulding23}. On the other hand, radio emissions can penetrate through dense columns of gas and dust, and current facilities can reach the required sensitivities. Still, AGN that exhibit a significant excess of \textit{non-thermal} radio emission above what would be expected from star-formation and thermal AGN contribution alone (defined as Radio-Loud; RL) are rare ($<10\%$ of the total AGN population, \citealt{kellerman89,vlba1p4}). 

Interestingly, a powerful, heavily obscured radio source ($L_\mathrm{Bol} \sim 10^{46}$\,erg\,s$^{-1}$) was discovered at $z\sim7$ (COS-87259, \citealt{Ends22}) -- and even this object posed more questions than it answered. COS-87259, first identified in the COSMOS field thanks to the broad bandwidth and depths accessed in the COSMOS survey, was recently spectroscopically confirmed at $z=6.8$ via [CII] detection in ALMA Band 6 observations \citep{ends23}. Bona fide evidence of the central engine in COS-87259 was discovered due to its bright radio emission. At $z\sim 7$, space density estimates of UV-bright sources are estimated to be 1/3000\,deg$^2$ \citep{shen20}, and for powerful RL AGN, 1/5000\,deg$^2$ \citep{rlspacedensity}-- yet this source was found in a HSC 1.5\,deg$^2$ survey. %While this could be ascribed to serendipity, the existence of a second source in this same 1.5 sq. deg. begins to unravel our understanding of the space density of these sources. 

Current UV-based absorption-corrected space density estimates imply that 10\% of the cosmic black hole growth in the Universe occurred by $z=6$ with a rapid build-up of growth occurring between $z=4$ and 2 \citep{shen20,matsuoka}. Increasing the number density of obscured sources above $z=6$ inspires several nuanced questions: Is there a significant reshaping of the gas distribution in AGN host galaxies that rapidly occurs between $z=7$ and 6? Are the UV bright AGN a much smaller tail of a larger AGN population -- and thus, our understanding of the number density estimates and accretion history of SMBHs over cosmic time needs to be overhauled? It is difficult to answer these questions when only one heavily obscured AGN at $z\sim7$ has been identified, i.e.,\ COS-87259. 
%Unfortunately, there is currently no one wavelength that is efficient in finding more of them. COS-87259 was selected via a HSC narrow-band survey to find epoch of re-ionization SF galaxies. The singular wavelength that hinted this source might harbor a buried QSO was the strong radio-emission detected via ancillary VLA data, and it was reported this source was RL. Even in the local Universe, RL QSOs are only a subset of the total QSO population ($<$10\%) and at $z=7$ have a measured space density of 1/5000\,deg$^2$ \citep{rlspacedensity}. %Thus the finding of a radio-loud, heavily obscured QSO in 1.5 sq degrees at z=6.8 is at the intersection of increasing improbability. % The detection of COS-87259 in a 1.5 sq deg. field can have three interpretations: 1) The space density of these sources is underestimated by over a factor of 2000, 2) A very rapid change in the gas properties of QSO host galaxies occurs from z=7 to z=6 to match directly measured obscured QSO fractions and the bolometric QSO luminosity function or 3) The discovery was entirely coincidental, with the assumption that surveying an additional 2000 sq. degs centered around COS-87259 would yield no other highly luminous QSO.

In this letter, we report the discovery of COSW-106725 in the COSMOS-Web field. This source was initially detected in the NIR (UVISTA + HST WFC3IR), radio (VLA + VLBA), and sub-mm (ALMA 343\,GHz continuum). During the April 2023 JWST Cycle~1 COSMOS-Web program observations, 4 NIRCam + 1 MIRI bands were imaged. Section~\ref{sec:obs} describes the observations of X-ray to sub-mm data of the source. Section~\ref{sec:results} reports the results from SED fitting and describes the derived AGN and galaxy properties. Section~\ref{sec:disc} compares the source to the only similar source on record and contextualizes these findings regarding high-z obscured AGN demographics. In Section~\ref{sec:disc}, we present the summary and conclusion. We use an $h = 0.7$, $\Omega_{m} = 0.3$, $\Omega_{\Lambda} = 0.7$ cosmology throughout this paper.

%We note this is only a sub-sample of the 19 photometric bands used in photo-z calculation.

\begin{deluxetable}{cc}[h]
\tablecaption{\label{tab:phot}Multi-wavelength ground- and space-based photometry for COSW-106725. All upper limits are at the 3$\sigma$ level.}
\tablecolumns{2}
\tablehead{  \colhead{Band}   &   \colhead{Flux ($\mu$Jy)}  }
\startdata
Subaru/HSC $g$      & $<$ 0.021    \\
Subaru/HSC $r$      & $<$ 0.034    \\
Subaru/HSC $i$      & $<$ 0.043    \\
HST/WFC3 \textit{F814W}  & $<$ 0.048    \\
Subaru/HSC $z$      & $<$ 0.063    \\
Subaru/HSC $y$      & $<$ 0.093    \\
JWST/NIRCam \textit{F115W} & 0.092 $\pm$ 0.002   \\
JWST/NIRCam \textit{F150W} & 0.21 $\pm$ 0.009    \\
HST/WFC3 \textit{F160W}  & 0.22 $\pm$ 0.011    \\
JWST/NIRCam \textit{F277W} & 1.0 $\pm$ 0.09      \\
\textit{Spitzer}/IRAC 3.6\,$\micron$ & 3.05 $\pm$ 0.4  \\
JWST/NIRCam \textit{F444W}      & 5.34 $\pm$ 0.05    \\
\textit{Spitzer}/IRAC 4.5\,$\micron$ & 5.5 $\pm$ 0.49  \\
\textit{Spitzer}/IRAC 5.8\,$\micron$ & 7.71 $\pm$ 0.57  \\
JWST/MIRI \textit{F770W}      & 11.0 $\pm$1.33   \\
\textit{Spitzer}/MIPS 24\,$\micron$  & 91.3$\pm$ 27.2   \\
\textit{Herschel}/PACS 100\,$\micron$       & $<$ 0.0012   \\
\textit{Herschel}/PACS 160\,$\micron$       & $<$ 0.0053   \\
\textit{Herschel}/SPIRE 250\,$\micron$      & $<$  0.021  \\
\textit{Herschel}/SPIRE 350\,$\micron$      & $<$  0.023  \\
\textit{Herschel}/SPIRE 500\,$\micron$       & $<$  0.012   \\
JCMT/SCUBA-2 850\,$\micron$     & $<$ 0.026   \\
ALMA 343\,GHz     & 2.5$\times$10$^{3}$ $\pm$ 5$\times$10$^{4}$   \\
VLA 3\,GHz        & 0.776$\times$10$^{3}$ $\pm$ 0.04$\times$10$^{4}$  \\
VLA 1.4\,GHz      & 1.78$\times$10$^{3}$ $\pm$ 0.15$\times$10$^{4}$  \\
MeerKAT 1.28\,GHz & 1.99$\times$10$^{3}$ $\pm$ 8.8    \\
GMRT 610\,MHz      & 3.43$\times$10$^{3}$ $\pm$ 1.7$\times$10$^{4}$  \\
VLITE 338\,MHz      & 6.63$\times$10$^{3}$ $\pm$ 1.1$\times$10$^{3}$  \\
VLA 324\,MHz      & 6.92$\times$10$^{3}$ $\pm$ 4.8$\times$10$^{4}$  \\
GMRT 325\,MHz      & 6.27$\times$10$^{3}$ $\pm$ 3.1$\times$10$^{4}$  \\
LOFAR 144\,MHz      & 8.91$\times$10$^{3}$ $\pm$ 1.9$\times$10$^{4}$  \\
\enddata
%\tablecomments{\textcolor{blue}{[note from TAH:  I added some of the telescope names (didn't do all of them cause some I don't know), but also happy to remove those and just leave the instrument names instead]}}
\end{deluxetable}

\section{Multi-wavelength Observations of COSW-106725}
\label{sec:obs}

The target was first erroneously classified over 10 years ago during a search for low-luminosity radio galaxies at cosmic noon within the COSMOS field (COSMOS-FRI-07, see \citealt{Chia09} for details). COSMOS is a deep, wide area, multi-wavelength survey centered on RA 10:00:30.12, Dec +2:12:38.80 \citep{Scov07}. Extensive observations of the field from almost all major space- and ground-based telescopes have accrued over the past 20 years \citep{laigle16,Weav22}. The initial basic selection criteria of \citet{Chia09} was based on the initial COSMOS multi-wavelength catalog \citep{cosmosog} and initial VLA 1.4 GHz observations \citep{cosmosvlaog}. This required the radio flux (at 1.4\,GHz) to be between 1 and 13\,mJy and the optical magnitude to be higher than i+ = 21 (Vega). Although COSW-106725 made the initial sample selection in the radio range, the source was erroneously associated with the combined optical detections of a bright star and a lower-z interloper within 2\arcsec\ of the radio coordinates. The initial NIR (CFHT) and MIR (Spitzer/IRAC) \citep{sanders07} fluxes were also highly uncertain due to poor spatial resolution and multiple interlopers. The limiting spatial resolution of the optical-MIR data and the dis-concordance between the radio and the source's optical properties were noted, and the nature of the object was left unknown. 

In the past ten years, deeper imaging and new wavelength coverage have been taken in the COSMOS field. In addition to deeper radio data, larger radio coverage, and a growing number of ALMA observations -- the central 0.54\,deg$^2$ of the COSMOS field was chosen for the largest JWST program scheduled for observations during the observatory’s first cycle in both sky coverage and total prime time allocation \citep[COSMOS-Web Survey, PID \#1727, PIs: Kartaltepe \& Casey;][]{cosmosweb}. COSMOS-Web consists of one large contiguous 0.54\,deg$^2$ NIRCam mosaic conducted in four filters, with additional MIRI imaging covering 0.18\,deg$^2$, and will be completed by January 2024. Within the current 0.27\,deg$^2$ covered, this combination of new data in the COSMOS field has lifted the veil of uncertainty around COSW-106725 -- and allowed us to identify the highest-redshift heavily obscured radio-loud AGN candidate to date. In the following sub-sections, we highlight the relevant observations conducted since the initial discovery of COSW-106725. 

\begin{figure}[h]
%\centering
\includegraphics[width=1.1\columnwidth]{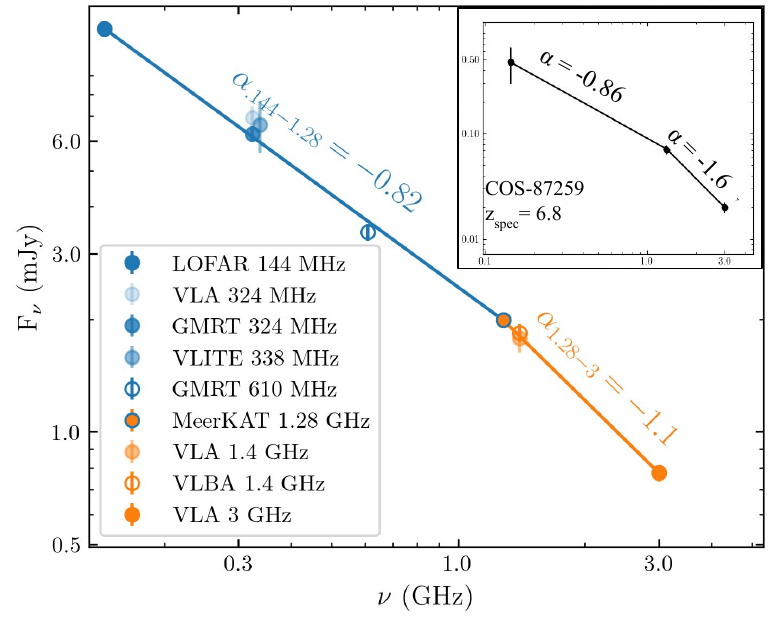}
	\caption{Radio SED: All fluxes and associated errors are listed in Table~\ref{tab:phot}. We measure the spectral slope between two sets of radio frequencies (blue line, orange line) and find significant spectral steepening indicative of high-$z$ RL AGN \citep{saxena18a,Ends22,broderick22}. In the upper-right corner inset, we show the radio SED for the z$_{spec}=6.8$ heavily obscured RL AGN from \citealt{Ends22} for reference.} 
\label{fig:radio_sed}
\end{figure}

\subsection{Radio}
The COSMOS field has been observed over a large range of radio wavelengths (144\,MHz--3\,GHz) via the Very Large Array (VLA), Very Long Baseline Array (VLBA), the Giant Metrewave Radio Telescope (GMRT), and the International Low-Frequency Array (LOFAR). COSW-106725 was strongly detected with  LOFAR HBA at 144MHz (8.51 $\pm$ 1.9\,mJy; DDT19\_002; PI: Vardoulaki), GMRT 325\,MHz (6.27 $\pm$ 0.480\,mJy) and VLA 324 MHz (6.93 $\pm$ 0.5\,mJy) \citep{gmrt324}, MeerKAT 1.28 GHz (1.99 $\pm$ 8.8$\times 10^{-3}$\,mJy) \citep{meerkat}, VLA 1.4\,GHz (1.78 $\pm$ 0.15\,mJy) \citep{cosmosvlaog}, VLBA + GBT 1.4\,GHz (1.84 $\pm$ 0.1\,mJy) \citep{vlba1p4}, and VLA 3\,GHz (0.776 $\pm$ 0.04\,mJy) \citep{vla3}. The physical extent of the VLA 3\, GHz detection deconvolved with the beam is $<$2.2\arcsec.

COSW-106725 is also detected as a compact source at the $\sim$5$\sigma$ level in all 3 epochs of the VLA Sky Survey \citep[VLASS;][]{vlass}.  The peak flux density averaged over the 3 VLASS epochs and measured from the quick-look image products is ~0.711 mJy/beam.  This measurement is consistent with the VLA 3 GHz measurement reported in Table \ref{tab:phot}.  We do not find any evidence for significant variability at 3 GHz given the typical 20\% flux scale uncertainty in VLASS quick-look data. To our knowledge, COSW-106725 is the highest redshift source detected in VLASS so far, surpassing the VLASS detection of a quasar at z $\sim$ 6.2 in \citealt{banados23}. Furthermore, there is a robust detection of COSW-106725 from the VLA Low-band Ionosphere and Transient Experiment (VLITE\footnote{\url{https://vlite.nrao.edu}}) which commensally records data at a center frequency of 338 MHz with nearly all VLA observations \citep{clarke2018,polensky}. COSW-106725 was detected across many individual observations with VLITE. The average total flux of the source is 6.63$\pm$1.05 mJy taking into account the 15\% flux uncertainties of VLITE.

In Figure~\ref{fig:radio_sed}, we plot the radio SED of COSW-106725. Using $\mathrm{S}_{\nu} \propto \nu^{\alpha}$, we measure the radio slope between 144\, MHz and 1.28\, GHz ($\alpha_\mathrm{.144-1.28} = -0.82$) and the radio slope between 1.32 GHz and 3 GHz ($\alpha_\mathrm{1.28-3} = -1.1$). This spectral steepening toward higher frequencies is consistent not only with the reported radio properties of COS-87259 but also with the behavior of many spectroscopically confirmed $z>4$ RL AGN \citep{miley2008,saxena2018b,Yamashita20,Drouart20,broderick22}. 

Finally, we compare the \textit{Spitzer} MIPS 24\,\micron\ and VLA 1.4\, GHz fluxes to assess the level of non-thermal AGN contribution to the radio emission. The observed 24\,\micron\ and 1.4\, GHz fluxes are tightly related for thermal sources (i.e non-RL AGN and star-forming galaxies). Using the parametrization in \cite{bonzini}, we measure the value of q$_{24\mathrm{obs}} = \mathrm{log}_{10}(f_{24\micron}/f_{1.4 \mathrm{GHz}}) = -1.1$, indicating the presence of powerful radio emission due to a kpc-scale jetted AGN or compact radio source vs thermal emission associated with radio-quiet AGN and/or star-formation.

\begin{figure*}[h!]
\centering
\includegraphics[width=1\textwidth]{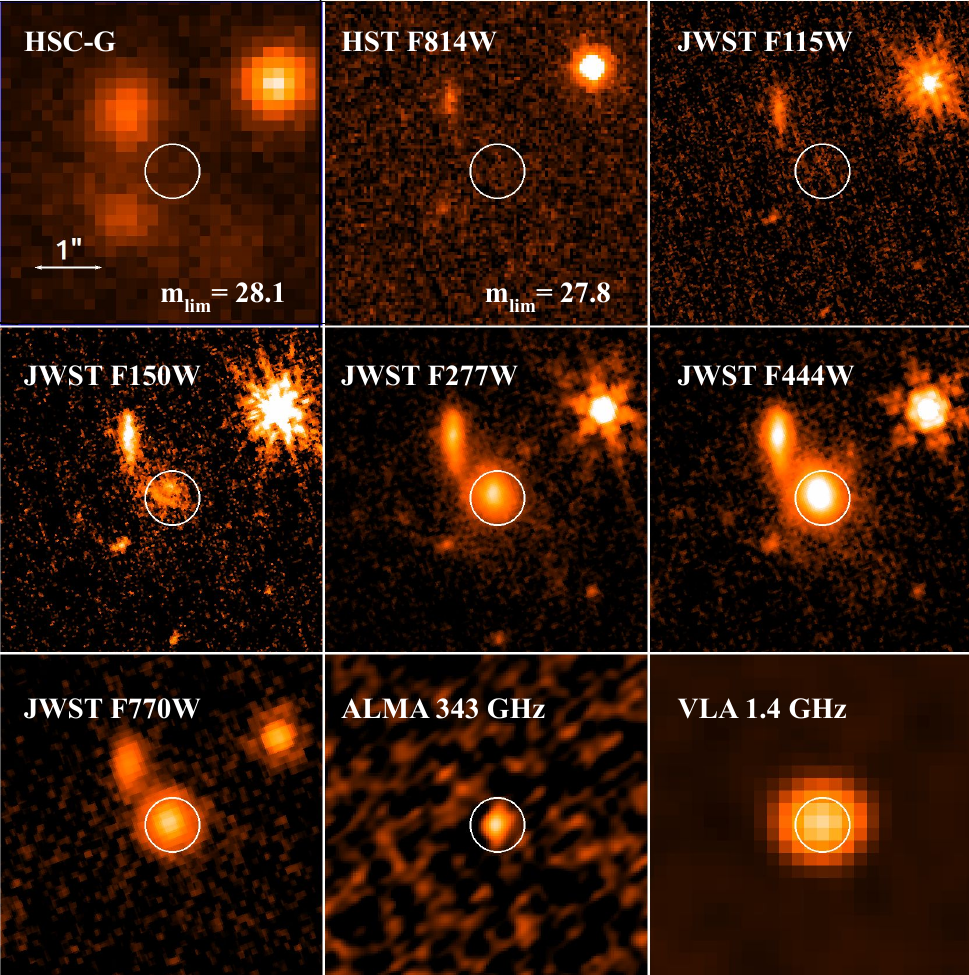}
	\caption{Selection of Postage Stamp Images of the Candidate $z\sim7.7$ RL Quasar, COSW-106725. Top row, from left to right: HSC-$g$, ACS \textit{F814W}, JWST \textit{F115W}, JWST \textit{F150W}, JWST \textit{F277W} and, JWST \textit{F444W}, JWST MIRI \textit{F770W}, ALMA 343\,GHz and VLA 1.4\,GHz. The ALMA extent is overlaid on each image (in white). The 3$\sigma$ upper-limits are reported for the non-detections. The upper left source in the UV/Optical/NIR images is a low-$z$ interloper \cite{Weav22}.} 
\label{fig:sub_props}
\end{figure*}

\subsection{ALMA}

COSW-106725 has a robust 5$\sigma$ detection ($F_{\mathrm{int}}= 2.5 \pm 0.5$ mJy) in $\sim$870\,\micron\ band continuum imaging via the A3COSMOS catalog \citep{a3cosmos}. The A3COSMOS catalog used the rich public Atacama Large Millimeter/Submillimeter Array (ALMA) archive to generate automated mining pipelines across the COSMOS field. We use the Gaussian fit flux via the ``blind'' pipeline. We note the ``prior''-fitting photometry catalog  
 yields an equivalent flux measurement (see \citet{a3cosmos} for details). 
 
\subsection{X-ray}\label{sec:xray}

The source was previously covered with the Chandra ACIS-I detector \citep[160\,ks,][]{chandralegacy} and the XMM-Newton PN, MOS1, and MOS2 detections \citep[30\,ks,][]{xmmcosmos}. This source is un-detected in the Chandra-Legacy survey of the COSMOS field and the XMM-COSMOS survey \citep{civano16,xmmcosmos}. We calculate the upper-limit 2--10\,keV flux in the 160\,ks combined event image using the \texttt{CIAO} tools function \texttt{aprates} \citep{ciaotools}, and find $F_\mathrm{2-10\,keV} < 2.3 \times 10^{-15}$ erg s$^{-1}$ cm$^{-2}$. In Section \ref{sec:disc}, we further discuss the X-ray upper-limits.

\begin{figure*}[t]
\centering
\includegraphics[trim= .1cm .1cm .2cm .2cm, scale=.8]{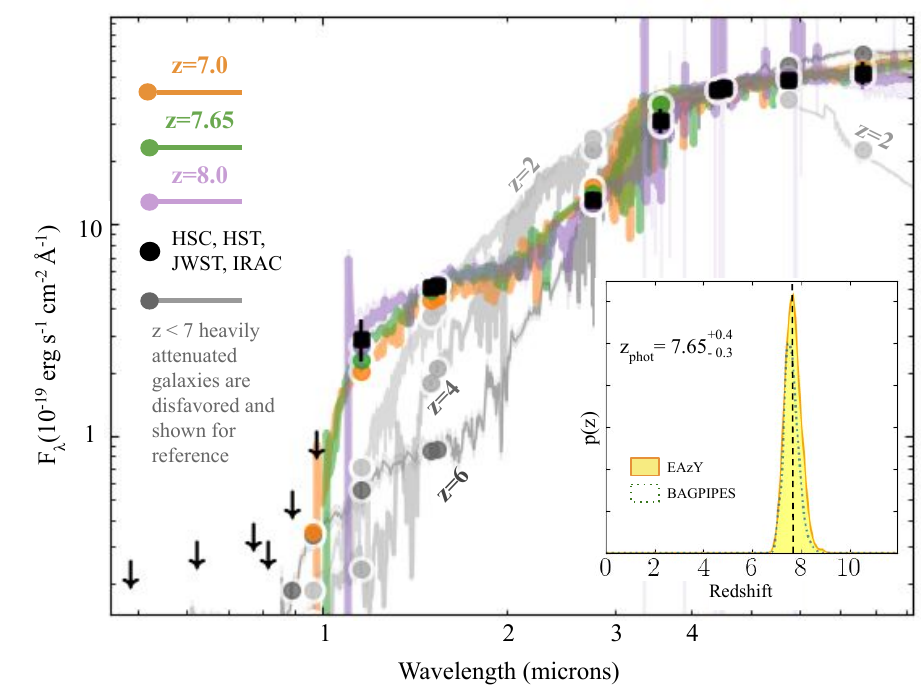}
\caption{Results from fitting the optical, NIR and MIR with {\emph{\texttt{EAzY}py}}. Non-detections with 27 mag upper limits: HSC $g$, HSC $r$, HSC $i$, HSC $z$, HST \textit{F814W}, HSC $y$. $>3 \sigma$ detections: JWST \textit{F115W}, JWST \textit{F150W}, HST \textit{F160W}, JWST \textit{F277W}, IRAC Channel 1, JWST \textit{F444W}, IRAC Channel 2, IRAC Channel 3, JWST MIRI 7.7\,\micron. The redshift is constrained to $z = 7.7^{+0.4}_{-0.3}$ fit with combinations of SSP template from \citep{bc03}. Inset: We show the p(z) via EAzY and BAGPIPES}
\label{fig:photoz}
\end{figure*}

\subsection{Additional Ground and Space-Based Optical/NIR/MIR Imaging} \label{sec:mips}
All optical upper limits are drawn from the ``classic'' COSMOS2020 catalog \citep{Weav22}. Included in COSMOS2020 is ultra-deep, broad-band photometry from the second public data release of the Hyper Suprime-Cam (HSC) Subaru Strategic Program comprising the $g$, $r$, $i$, $z$, and $y$ bands. COSW-106725 is undetected in all bands ($g$: mag$_{\mathrm{lim}}$ = 28.1, $r$: mag$_{\mathrm{lim}}$ = 27.8, $i$: mag$_{\mathrm{lim}}$ = 27.6, $z$: mag$_{\mathrm{lim}}$ = 27.2, and $y$: mag$_{\mathrm{lim}}$ = 26.5). HST/ACS \textit{F814W} high-resolution photometry is also included, and the object remains undetected (mag$_{\mathrm{lim}}$ = 27.8). 

A search of COSW-106725's radio coordinates in MAST serendipitously finds a WFC3IR F160W image of another source covered in an unrelated HST campaign (PI: Conselice, Cycle 24, GO:14721). Using Source Extractor \citep{sourceex} on the MAST reduced image, we measure a 1\arcsec\ aperture \textit{F160W} flux that agrees with the JWST \textit{F150W} flux. This source is also detected in all four Spitzer IRAC bands and MIPS 24\,\micron. We use the source locations in the JWST NIRCam \textit{F277W} and radio bands to deblend the Spitzer photometry and find excellent photometric agreement with JWST NIRCam \textit{F444W} and IRAC Ch~2 (see \citealt{jin18} for details). 

The positional accuracy of the radio, ALMA, and JWST emission are all within 1\arcsec. All fluxes and upper limits are listed in Table~\ref{tab:phot}. 

% Erini - finish this sentence and but it in the para above: Using the detection image in the JWST NIRCam F277W filter, we de-blend the IRAC fluxes using

\begin{figure*}[htb]
    \centering
    \includegraphics[width=1\textwidth, trim=0 5mm 0 0]{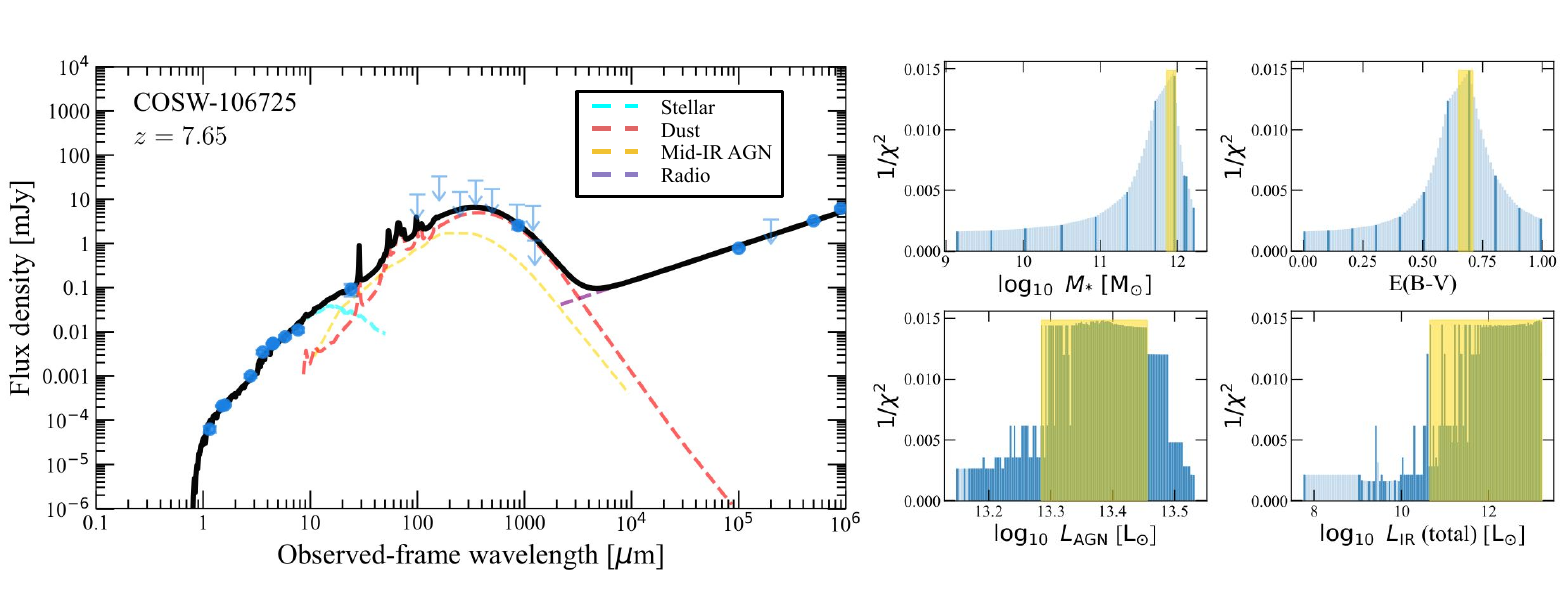}%\\
    \caption{\textit{Left Panel:} Optical-IR-radio SED fitting with BC03 stellar \citep{bc03}, mid-IR AGN \citep{mullaney11}, Draine \& Li dust \citep{draineli07} and power-law radio templates (using the MICHI2 code; \citep{liu21}). The black line indicates the composite best-fit model and the blue symbols are photometric data points, with upper limits shown as downward arrows. The stellar, mid-IR AGN, dust, and radio components are indicated by the cyan, yellow, red, and magenta dashed lines, respectively. 
    \textit{Right panels:} The 1/$\chi^2$ distributions from the fitting for the four parameters: stellar mass, dust attenuation $E(B-V)$, AGN component's luminosity integrated over 10-1000~$\mu$m, and dust component's luminosity integrated over 8-1000~$\mu$m. The yellow highlighted regions correspond to the 95\% confidence intervals.}
    \label{fig:Opt-IR-radio SED}
\end{figure*}

\subsection{JWST NIRCam+MIRI Imaging}

This object is in the Cycle~1 JWST COSMOS-Web field (GO \#1727, PIs: Kartaltepe \& Casey, \citealt{cosmosweb}), with observations available in four NIRCam wide-band filters: \textit{F115W}, \textit{F150W}, \textit{F277W}, and \textit{F444W}, and one MIRI wide-band filter: \textit{F770W}.  Forthcoming papers will comprehensively describe the complete data reduction process (COSMOS-Web NIRCam; M.\ Franco et al., COSMOS-Web MIRI; S.\ Harish et al.), but we briefly outline the procedures here. Upon retrieval of the uncalibrated NIRCam images from the STScI MAST Archive, we reduced the data utilizing the JWST Calibration Pipeline \citep{bushouse}. Custom modifications were incorporated, such as mitigating 1/f noise and subtracting low-level background, following the precedent set by other JWST studies \citep[e.g.,][]{bagely23}. All reference files, including in-flight data, represented the latest calibrations available during our observations. The final mosaics were generated during Stage 3 of the pipeline, varying only in resolution, with pixel sizes of 0.03\arcsec/pixel and 0.06\arcsec/pixel. Unless otherwise specified, we will refer to the 0.06\arcsec/pixel resolution mosaic hereafter. The JWST mosaics were aligned to a version of the COSMOS F814W mosaic \citep{Koekemoer2007} that had been astrometrically aligned to Gaia DR3, with the F814W mosaic subsequently used as a reference catalog for all the JWST imaging \citep{Koekemoer2007}. The median offset between the \textit{F814W} mosaic and the COSMOS-Web NIRCam mosaic is less than 5\,mas. 

The MIRI \textit{F770W} observations were also reduced using the JWST Calibration pipeline and with the additional background subtraction step to mitigate instrumental effects. The \textit{F770W} mosaic was then resampled to an output grid corresponding to 0.06\arcsec/pixel and aligned with the HST ACS \textit{F814W} imaging. We perform source detection and measure the multi-wavelength aperture photometry of the COSMOS-Web data using Source Extractor V~2.86 (SE, \citealt{sourceex}). We use 1\arcsec\ apertures and apply a detection threshold corresponding to a signal-to-noise ratio (S/N) of 3.

\section{Results}
\label{sec:results}

\subsection{Photo-$z$ Estimate via Optical/NIR/MIR Photometry}\label{sec:photz}

With the photometry listed in Table~\ref{tab:phot}, we first run \texttt{EAzY}, a template-based SED fitting code \citep{Bram08}. \texttt{EAzY} generates a photo-$z$ probability density function (PDF) via $\chi^{2}$ minimization using linear combinations of pre-defined templates. We use the standard 12 template FSPS set included in \texttt{EAzY} (tweak\_fsps\_QSF\_12\_v3) and the 6 additional templates from \citealt{larson22}. In conjunction with the deep ground-based data and the unprecedented resolution and sensitivity of JWST -- we perform robust SED fitting on the source and find a singular solution photo-$z$ estimate of $z_\mathrm{phot} = 7.7^{+0.4}_{-0.3}$ with reduced $\chi^{2} = 0.3$. 

The Balmer break spectral region is well sampled with IRAC+JWST observations, and the Lyman break is sampled via deep HSC/HST+JWST observations. The detection level in \textit{F115W} places a strict $z < 8$ constraint. In Figure~\ref{fig:photoz}, we also show (gray lines) the fits to templates of heavily dust-obscured star-forming galaxies at $z < 7$. The HST \textit{F160W} + JWST \textit{F150W}/\textit{F277W} detections heavily disfavor any templates with $2 < z < 7$ while the MIRI \textit{F770W} detection solidly rules out the $z \ge 2$ templates. The IRAC data used in the fit well samples the data as is evidenced by the similar fluxes in IRAC Ch~2 (4.5\,\micron) and JWST \textit{F444W} (4.4\,\micron). In addition to the fit photo-$z$, the ancillary observations of this source robustly constrain the redshift to within $z=7$--8. 

We also independently measure the photo-z using \texttt{BAGPIPES} using a delayed-tau star-formation history ($\log (M_{*}/\mathrm{M}_{\odot}) \sim 6$--13, Z$\sim$0.001--2.5, $\tau \sim 0.1$--5\, Gyr, Age $\sim 0--100\%$ $t_{\mathrm{H}}$), constant starburst (Age $\sim$ 1--100\, Myr), nebular emission ($\log U \sim -4$ to $-1$), flexible dust attenuation law ($A_{\mathrm{V}} \sim 0$--3, slope allowed to vary with a Gaussian prior centered on an SMC dust law), and redshift ($z\sim 0$--12). We find a consistent photo-$z$ ($z=7.5 \pm 0.35$, $\chi^{2} = 0.27$ ), $A_\mathrm{V}\sim 2$, and a $M_{*} = 2.8$--5.4 $\times 10^{11}$ M$_{\odot}$. In Figure \ref{fig:photoz}, we overlay the \texttt{BAGPIPES} p(z) in the lower-left inset.

\subsection{SED Decomposition at Best-Fit Photo-z}
\label{sec:seddecomp}
%%% 20230703, Daizhong's editing, feel free to edit/delete anything! %%%

Using the photo-$z$ derived via \texttt{EAzY}, we then fit the global optical--IR--radio SED with a composite of SED components accounting for stars, mid-IR AGN, dust, and radio emission to produce tighter constraints on the stellar mass and infer the AGN bolometric luminosity (Fig.~\ref{fig:Opt-IR-radio SED}). We use the MICHI2 code\footnote{\url{https://github.com/1054/Crab.Toolkit.michi2}; \citep{liu21}} to fit multiple SED components simultaneously: a) the BC03 \citep{bc03} synthesized stellar templates (with a constant star formation history and \citealt{calzetti} attenuation law), b) the low-redshift observationally-constructed mid-IR AGN templates \citep{mullaney11}, c) the widely-used Draine \& Li dust models \citep{draineli07}, and d) a power-law radio component with a spectral index 0.8, consistent for most radio-loud AGN \citep{vla3}. 

The best-fit SED shows a strong contribution from the AGN in the mid-IR, dominating the 20--200\,\micron\ emission. The 1/$\chi^2$ distributions representing the parameter probabilities are shown in the right panels of Fig.~\ref{fig:Opt-IR-radio SED}. Taking into account the redshift posterior distribution in the error propagation from \texttt{EAzY}, we find a well constrained stellar mass $\sim$$10^{11.92 \pm 0.5} \ \mathrm{M_{\odot}}$, dust attenuation of $E(B-V) \sim 0.68 \pm 0.08$, and a loosely constrained dust infrared luminosity $\sim$$10^{12} \ \mathrm{L_{\odot}}$ (which has the AGN contribution subtracted). 

The fitted AGN luminosity integrated over 10--1000\,\micron\ is $\sim$1--3$ \times 10^{13}\, \mathrm{L_{\odot}}$, corresponding to an AGN bolometric luminosity of $\sim$4--12$ \times 10^{46}$\,erg s$^{-1}$ via the bolometric correction provided in \citet{delve14}. The bolometric luminosity of the source, coupled with the lack of any point source in the NIR images and lack of detection in the Chandra-Legacy 160\, ks survey, allows us to infer the level of obscuration of the AGN to be $N_{H} > 10^{23}$\,cm$^{-2}$. Given that this quasar is heavily obscured in the optical, we do not include a rest-frame UV-optical quasar template in our fitting. 

% X-ray to bolometric conversion (vice versa) so can convert from LbolSED to LxSED and then compare to Lxupperlimit. 

 % via re-estimating the lower limit of the 2-10 keV X-ray flux of the source using the L$_{\mathrm{Bol}}$ found via the multi-band SED fitting ($\ge$ 4-12 $\times 10^{46}$ erg/s) and the X-ray bolometric correction. 

Next, we compare the above SED-derived $L_{\mathrm{Bol}}$ to the $L_{\mathrm{Bol}}$ estimated from the X-ray upper limit. We apply the correction provided in \citet{duras20} to estimate the hard-band X-ray luminosity from the bolometric luminosity derived via SED fitting and calculate: $L_{2-10 \mathrm{keV,SED}} = 2$--10$ \times 10^{44}$\,erg s$^{-1}$. We then calculate the X-ray 2--10\,keV luminosity using the X-ray flux upper-limit derived in Section \ref{sec:xray} and the photo-$z$ estimated from Section~\ref{sec:photz}, and find $L_{2-10 \mathrm{keV,X-ray}} < 1.5\times10^{45}$\,erg s$^{-1}$. Thus, assuming this object is at $z\sim7.7$, the bolometric luminosity derived from the optical--sub--mm SED fit agrees with the X-ray-based upper limit estimate.

%At this redshift, the observed-frame 2-10 keV flux of CT QSO should not differ much from that of an unobscured QSO (rest frame 16 - 80 keV) \cite{gilli07}. Using the photo-z estimate derived from the template fitting code \texttt{EAzY}, we estimate the 2-10 keV X-ray flux as $2.1 \times 10^{-15}$ erg/s/cm$^{2}$ assuming an intrinsic of $\Gamma = 1.8$ with an absorbed power law (N$_{\mathrm{H}}$ = 10$^{23}$ cm$^{-2}$). This flux is in agreement with the upper-limit derived in Section \ref{sec:xray} (F$_{2-10 kev} < 2.3 \times 10^{15}$ erg s$^{-1}$ cm$^{-2}$). T

%Using z=7.65, we measure the 1.4 GHz radio power. The radio power of the source ($1.2 \times 10^{34}$ erg/s/Hz) can be used to infer an L$_{Bol} = 5.1 \times 10^{46}$ erg/s using the relation in [need proper citation]. Thus, the bolometric luminosity derived from optical-sub-mm SED fit is in agreement with the radio-based estimate. 

%%%

\section{Discussion and Conclusions}\label{sec:disc}

Assuming Eddington accretion, $\lambda_{\mathrm{Edd}} = 1$, we provide a lower limit to the black-hole mass of COSW-106725. Following the canonical Eddington luminosity relationship using $L_{\mathrm{Bol}}$ = 5.1 $\times 10^{46}$\,erg s$^{-1}$, we find $M_{\mathrm{BH}} \geq 6.4 \times 10^8$\,M$_{\odot}$. While this number is only a lower limit, we can calculate whether COSW-106725 is potentially more massive than expected by comparing the $M_{*}$ derived from the SED fit in Section~\ref{sec:seddecomp} to the $M_{*}$ derived from local $M_{\mathrm{BH}}$ vs.\ $M_{*}$ scaling relations. Using Equation 8 from \citealt{Ding2020}, we find that the comparable stellar mass for this black hole mass should be $M_{*}$ = 3.69 $\times 10^{11}$\,M$_{\odot}$. Due to the $M_{\mathrm{BH}}$ being a lower limit, the scaling relation derived $M_{*}$ is also a lower limit and is below the SED fit derived $M_{*}$ value (8.3 $\times 10^{11}$\, M$_{\odot}$). Thus, our estimated $M_{\mathrm{BH}}$ does not indicate an over-massive BH concerning its host galaxy. 

In summary, we report the discovery of COSW-106725 in the COSMOS-Web field. The coincident radio/sub-mm/JWST observations of the source provide a robust estimate of $z_{\mathrm{phot}}$ = 7.7. This source is first detected in the rest-frame optical via JWST \textit{F115W} and remains undetected in deep space- and ground-based 0.4--1\,\micron\ imaging. Due to the high-inferred $L_{\mathrm{Bol}}$ = 5.1 $\times 10^{46}$\,erg s$^{-1}$ and lack of significant AGN emission in the rest-frame optical/NIR/X-ray, the source is inferred to be an intrinsically powerful, and heavily obscured ($N_{\mathrm{H}} > 10^{23}$\,cm$^{-2}$) AGN. Thus leading to its classification as a Type 2 AGN candidate. The detection of this source (COSW-106725) and COS-87259 \citep{ends23} within the epoch of $z=6.8$--8 in a 1.5\,$\deg^2$ field hints that the space density of luminous, radio-loud AGN at these epochs may be underestimated by over a factor of 2000. Even in the local Universe, radio-loud AGN are only a subset of the total AGN population ($< 10\%)$ and at $z=7$ up to now have had a measured space density of 1/5000\,$\deg^2$ \citep{rlspacedensity}. Thus, the discovery of two radio-loud, heavily obscured AGN within 1.5\,$\deg^2$ at $z\sim7$ is at the intersection of increasing improbability \citep{rlspacedensity}. 

For there to be more AGN in the Epoch of Reionization than predicted via extrapolation of luminosity functions at lower redshifts, a very rapid change in the gas properties of AGN host galaxies must occur \citep{vito2018}. Selecting heavily obscured sources at high redshift remains challenging even with JWST, and answering the nuanced questions surrounding early BH formation and growth with sparse data sets is challenging. Thus, combining JWST imaging with deep radio data can potentially revolutionize our understanding of powerful, obscured sources at cosmic dawn by enabling their efficient selection.

\begin{acknowledgments}
Acknowledgments: We thank R. Gilli for their incredibly useful discussion. We also thank the anonymous referees for their thoughtful insight and important contributions to this work. ELL and TAH are supported by appointment to the NASA Postdoctoral Program (NPP) at NASA Goddard Space Flight Center, administered by Oak Ridge Associated Universities under contract with NASA. The National Radio Astronomy Observatory is a facility of the National Science Foundation operated under cooperative agreement by Associated Universities, Inc. Basic research in radio astronomy at the U. S. Naval Research Laboratory is supported by 6.1 Base funding. Construction and installation of VLITE was supported by the NRL Sustainment Restoration and Maintenance fund. ALMA is a partnership of ESO (representing its member states), NSF (USA) and NINS (Japan), together with NRC (Canada), MOST and ASIAA (Taiwan), and KASI (Republic of Korea), in cooperation with the Republic of Chile. The Joint ALMA Observatory is operated by ESO, AUI/NRAO and NAOJ. Support for this work was provided by NASA through grant JWST-GO-01727 and HST-AR-15802 awarded by the Space Telescope Science Institute, which is operated by the Association of Universities for Research in Astronomy, Inc., under NASA contract NAS 5-26555. Some of the data presented in this paper were obtained from the Mikulski Archive for Space Telescopes (MAST) at the Space Telescope Science Institute. The specific observations analyzed can be accessed via \dataset[DOI]{https://doi.org/10.17909/ym93-d513}. %This paper employs a list of Chandra datasets, obtained by the Chandra X-ray Observatory, contained in~\dataset[DOI: X]{https://doi.org/X}.

\software{pandas \citep{pandas}, scipy \citep{scipy}, ipython \citep{ipython}, matplotlib \citep{matplotlib}, BAGPIPES \citep{bagpipes}, astropy \citep{astropy}, EAzY \citep{Bram08}}.

\end{acknowledgments}

\bibliography{v2_responses_incorporated_j10000104_photz_paper}{}
\bibliographystyle{aasjournal}

%% This command is needed to show the entire author+affiliation list when
%% the collaboration and author truncation commands are used.  It has to
%% go at the end of the manuscript.
%\allauthors

%% Include this line if you are using the \added, \replaced, \deleted
%% commands to see a summary list of all changes at the end of the article.
%\listofchanges

\end{document}